\documentclass[runningheads]{llncs}
\usepackage[T1]{fontenc}
\usepackage{graphicx}
\usepackage{hyperref} 
\usepackage{float}

\usepackage{microtype}

\usepackage{enumitem}
\setlist{nosep,leftmargin=*}



\makeatletter
\let\oldthebibliography\thebibliography
\renewcommand{\thebibliography}[1]{%
  \oldthebibliography{#1}%
  \setlength{\parskip}{0pt}%
  \setlength{\itemsep}{0pt plus 0.3ex}%
}
\makeatother

\begin{document}
\title{Learning Hanzi Character Through VR-Based Mortise-Tenon}
%
%
\author{
  Conglin Ma\inst{1} \and
  Jiatong Li\inst{1} \and
  Sen-Zhe Xu\inst{2} \and
  Ju Dai\inst{3} \and
  Jie Liu\inst{1} \and
  Feng Zhou\inst{1}
}
 
\institute{
  North China University of Technology, Beijing, China \\
  \and
  University of Science and Technology Beijing, China \\
  \and
  Peng Cheng Laboratory, China \\
}

%
\maketitle              
\vspace{-10pt} 
\begin{center}
 conglin04@mail.ncut.edu.cn, lijiatong@mail.ncut.edu.cn, senzhe@ustb.edu.cn, daij@pcl.ac.cn, liujxxxy@126.com, zhoufeng@ncut.edu.cn
\end{center}
\begin{abstract}
This paper introduces a novel VR-based system that redefines the acquisition of Hanzi character literacy by integrating traditional mortise-tenon joinery principles (HVRMT).Addressing the challenge of abstract character memorization in digital learning, our system deconstructs Hanzi components into interactive ``structural radicals'' akin to wooden joint modules.  Leveraging PICO's 6DoF spatial tracking and LLM's morphological analysis, learners assemble stroke sequences with haptic feedback simulating wood-to-wood friction. Our system also supports multiplayer online experiences, enhancing engagement and memory retention while preserving intangible cultural heritage. This innovative approach not only enhances engagement and memory retention but also reconstructs the craft wisdom embedded in Chinese writing systems, offering new pathways for preserving intangible cultural heritage in digital ecosystems. For the demo, please refer to \href{https://youtu.be/oUwfFTRpFyo}{this link}.

\end{abstract}

\section{Introduction}

In the field of digital education, the acquisition of Hanzi character literacy represents a unique and intricate learning task. Previous research has made substantial contributions to this domain, with traditional teaching methods providing systematic instruction in reading and writing \cite{vogel2022interactive,hong2013hanzi}. Nevertheless, these approaches are typically characterized by a “disembodied” nature, in which learners have limited opportunities for hands-on engagement or interactive exploration. As a result, it can be challenging to meaningfully connect the process of character learning with real-world contexts and cultural backgrounds, potentially leading to difficulties in retention and in developing a deep structural understanding of Hanzi \cite{hsiao2015influence}.
Currently, Chinese character education remains mired in traditional models. While methods such as memory reinforcement and emotional connection can improve short-term memorization, they remain constrained by the limitations of flat, book-based training and struggle to showcase the three-dimensional nature of Chinese characters as cultural vehicles. Some studies have attempted to incorporate gamification, such as the "Kanji Industry" system, which simulates an industrial assembly line through component assembly. However, these systems remain focused on the static display of Chinese characters, failing to establish a dynamic connection between their structure and traditional culture.
In recent years, virtual learning environments have emerged as promising tools to address some of these challenges. These platforms have introduced elements of interactivity and immersion, enriching the learning process beyond the static delivery of content. However, despite these advances, existing systems often face limitations in fully engaging learners with the cultural heritage and structural complexity embedded within Hanzi characters \cite{wang1998study}. For example, immersive, hands-on experiences—which have the potential to enhance spatial cognition and cultural understanding—are still relatively uncommon in current implementations \cite{hsiao2015influence}.

\begin{figure}[tp!]
    \centering
    \includegraphics[width=1\linewidth]{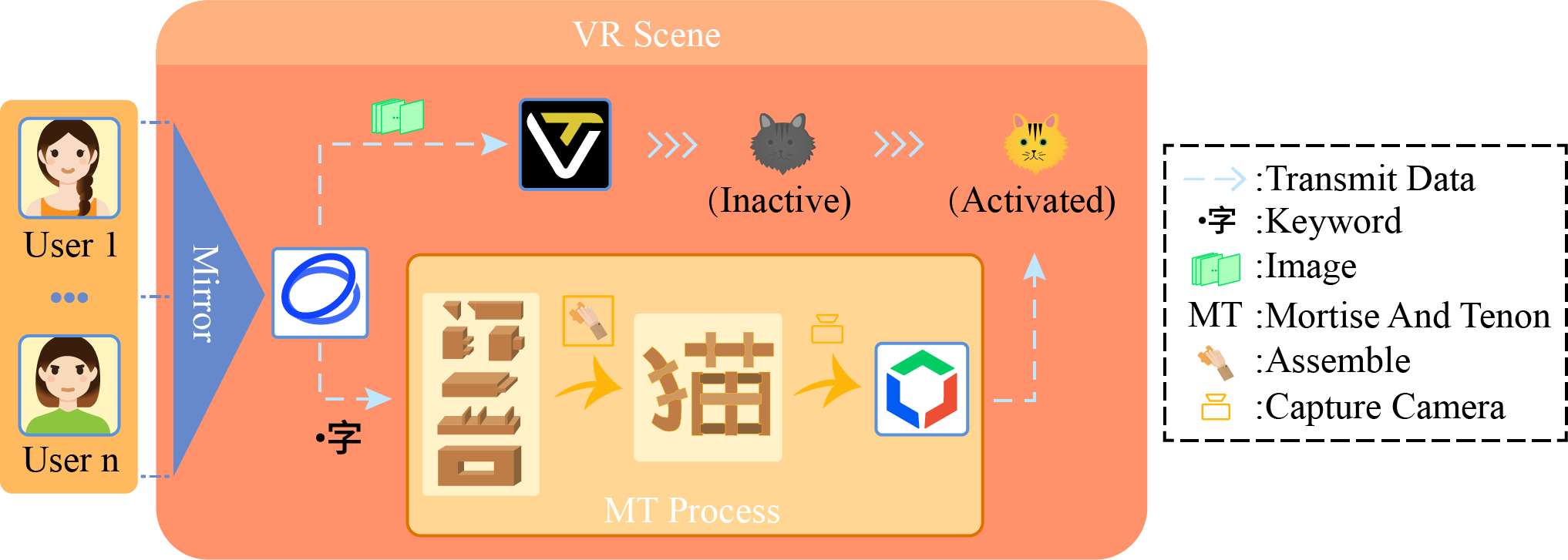}
    \caption{Overview of our HVRMT application.}
    \label{fig:enter-label}
\end{figure}

To further advance this field, the present study proposes a novel VR-based Hanzi learning system inspired by the ancient principles of mortise–tenon joinery. As illustrated in Figure \ref{fig:enter-label}, the system reimagines abstract Hanzi components as interactive “structural radicals,” analogous to the precise and functional wooden joint modules used in traditional Chinese carpentry. This design draws on the enduring craftsmanship of mortise–tenon structures, offering learners a tangible metaphor that bridges the gap between abstract character memorization and embodied, interactive learning.

Unlike traditional learning methods, the mortise–tenon joinery requires learners to carefully consider the shape and connection points of each component while assembling the mortise and tenon joints. This deeply engaging and interactive approach deepens their memory of the structure and form of Chinese characters. Furthermore, through engaging with the mortise and tenon joints of Chinese characters, learners not only learn about Chinese characters but also appreciate the charm and wisdom of this ancient craft.

Within this framework, learners physically assemble stroke sequences in a manner akin to architectural construction. Leveraging PICO’s advanced 6DoF spatial tracking technology, the system enables users to manipulate virtual components through natural hand movements, while a large language model (LLM)–driven morphological analysis module provides accurate interpretation and guidance. To enhance immersion, haptic feedback simulates the tactile sensation of wood-to-wood contact during component assembly, creating a multisensory learning environment that promotes engagement, improves memory retention, and deepens understanding of Hanzi’s structural logic. Furthermore, the system incorporates a virtual community feature, enabling multi-user collaboration and cultural exchange, thereby transforming character literacy learning into a socially enriched experience of cultural heritage transmission.

\section{Related Work}

Digital technologies are transforming the ways in which education and cultural heritage are preserved, and the emergence of LLMs has provided new avenues in this domain. A growing body of research is emerging in this area \cite{bu2025investigation,rao2024formationcreator}, and below we will focus on some methods that are directly related to our work.

\textbf{LLM Application in Educational Technology}: Recent applications of LLM in education have shown their potential in delivering personalized feedback and adaptive learning strategies. For example, the LEAP platform by Steinert et al. (2024) employs LLMs to generate formative feedback that supports self-regulated learning through text-based prompts such as self-explanation cues and motivational scaffolds~\cite{steinert2024leap}. However, these systems remain limited to textual interactions and metacognitive guidance, lacking tangible or interactive outcomes. In contrast, our system applies LLMs to a distinct functional role—understanding learner speech, performing morphological analysis, and generating corresponding 2D/3D interactive models. This bridges semantic comprehension with embodied interaction, enabling learners to construct and explore Chinese characters through immersive, model-driven experiences in VR.

\textbf{Cultural Heritage Preservation through Digital Interaction}: Many studies in human–computer interaction leverage digital technologies to revitalize traditional skills. For instance, Lee (2019) used AR to teach the three-dimensional structure and projection principles of mortise–tenon joints~\cite{lee2019mortisetenonAR}, showing that component-based visual teaching can enhance spatial cognition. However, this work focuses on process understanding and does not establish a semantic mapping between mortise–tenon structures and Chinese characters. Our study reconstructs these traditional principles as an interactive literacy module, systematically translating the idea of “using wood to build characters” into a virtual learning environment. Drawing on Yilmaz’s (2022) concept of cloud-based collaborative writing communities~\cite{yilmaz2022virtualwriting}, we extend this framework into a VR context by enabling multi-user interaction with mortise–tenon components, enhancing group participation and shared knowledge construction.

\section{System Design}
\vspace{-1mm}
Our HVRMT is founded on the harmonious fusion of traditional mortise-tenon and Hanzi character literacy acquisition. The following provides a detailed depiction of the system design:

\textbf{Core Concept}: The system maps Hanzi strokes to mortise-tenon components, transforming abstract characters into tangible, interactive modules. Learners assemble Hanzi radicals as if constructing wooden frameworks, bridging language learning with traditional craftsmanship.

\textbf{Technical Framework}: 1. Leverages PICO’s 6DoF spatial tracking to capture voice and movements, converting speech to text and extracting core characters. 2. Utilizes LLM for morphological analysis and guidance. Based on user input, it generates 2D images and 3D models of corresponding objects. 3. Provides mortise-tenon parts in the virtual space. A recognition camera captures the assembly process, verifies the formed character against the extracted core character, and activates the 3D model upon matching. 4. Utilizes the features of the mirror plugin to achieve synchronization and interaction among multiple users in VR scenes. Users use the VR thumbstick to move and turn, the trigger to complete UI interactions and other relevant operations, and the grip to pick up parts.

\section{Experiments} 

\begin{figure}
    \centering
    \includegraphics[width=1\linewidth]{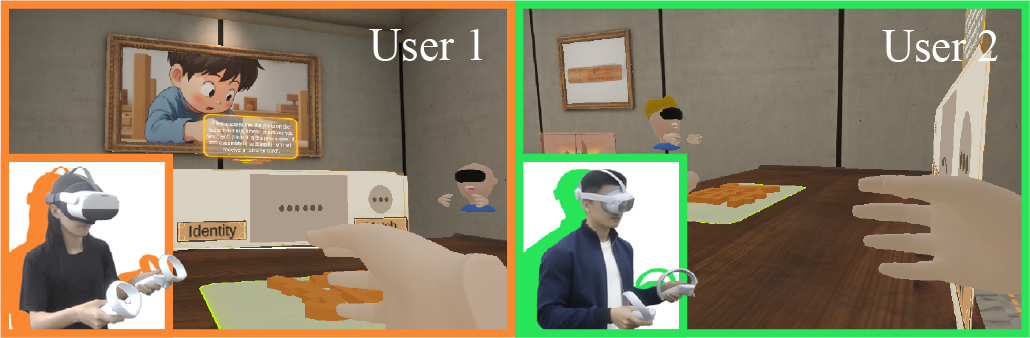}
    \vspace{-15pt}
    \caption{Two users scene design.}
    \label{fig:2}
\end{figure}

\textbf{Multi User System Design}: When multiple users wish to experience the system together,the system sets the first logged in user as the homeowner role. Subsequently, other users can utilize the system's built-in network multicast receiving function to enter the same scene as clients, thereby achieving a collaborative interaction experience among multiple users, as shown in Figure \ref{fig:2}.

\begin{figure}
    \centering
    \includegraphics[width=1\linewidth]{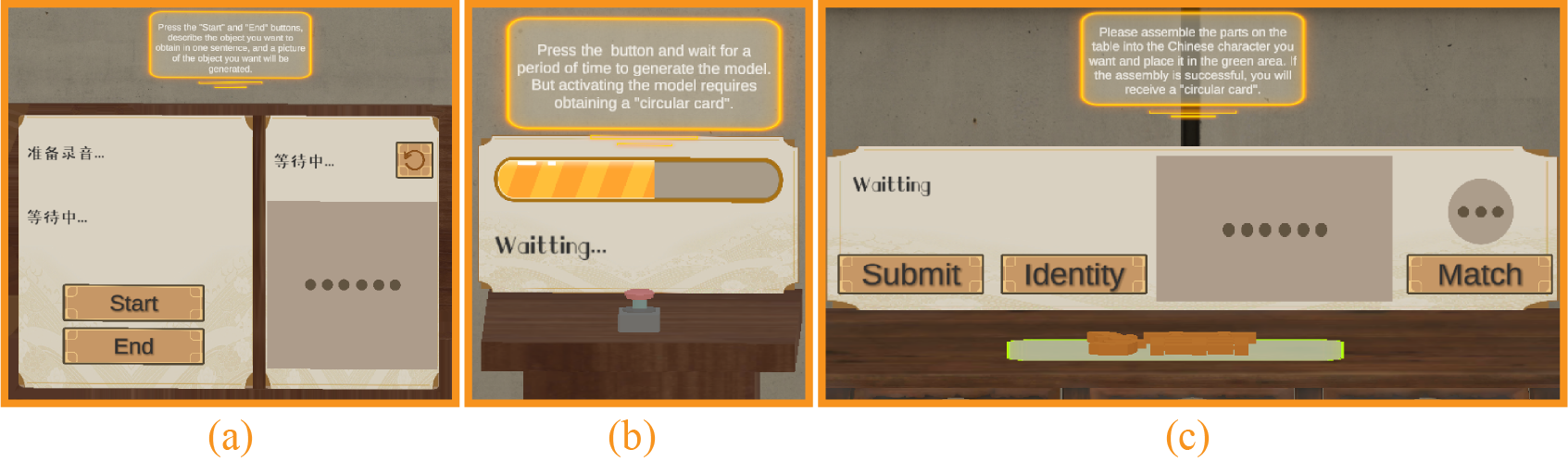}
    \vspace{-15pt}
    \caption{System area design diagram.}
    \label{fig:3}
\end{figure}

\textbf{User Interaction and Data Flow}: The virtual space is divided into three functional areas: 1. Speech Area (a): Speech recognition, keyword extraction, and image generation. 2. Model Area (b): 3D modeling and display. 3. Character Area (c): mortise-tenon assembly and OCR, as shown in Figure \ref{fig:3}.

Data flows from voice input to text conversion, core character extraction, image generation, and 3D model creation. Users assemble mortise-tenon parts in the virtual environment. The system captures the assembly, recognizes the character, compares it with the extracted core character, and activates the 3D model if they match.

\textbf{Experimental Procedure}: 
The user first logs in to the scene, then enters the voice area and describes the object they want to generate. For example, say "a cute cat". The system understands voice and text based on ChatGLM \cite{glm2024chatglm}. The Prompt is constructed as follows:
\textit{\textbackslash{}{"}model{"}:"glm-4-flash"\textbackslash{},  
\textbackslash{}{"}messages"\textbackslash{}:[\textbackslash{}{"}role"\textbackslash{}:"user"\textbackslash{},\textbackslash{}{"}content"\textbackslash{}:" 
+ text + ", extract the main object described in this sentence, ignore color and other modifiers, and require the result to be one character\textbackslash{}{"}]"\textbackslash{}"]}.
Then construct
\textit{\textbackslash{}{"}model{"}:"cogView-4-250304"\textbackslash{}, \textbackslash{}{"}prompt"\textbackslash{}:" + text + ", simple background, no complex environment, solid color background, clear subject"\textbackslash{}, \textbackslash{}{"}size"\textbackslash{}:"512x512"\textbackslash{}} to generate the corresponding image data.

Afterwards, the user can choose to enter the model area or the parts splicing area. In the model area, after the user clicks the "Generate" button, the system will call the image data generated by the voice module and use it as a basis to start the model generation process. The generated model is initially in the "unactivated" state and does not have interactive capabilities. After obtaining the round card in the subsequent stage, it can be activated and interactive functions can be realized. Use the Tripo interface to trigger the image generation model process by setting imagePath and calling Image\_to\_Model\_func(). After the generation is complete, listen to the OnDownloadComplete callback to obtain the download link (URL) of the model, and then use the GltfAsset component to load the model and display it in the scene.

\begin{figure}[h]
    \centering
    \includegraphics[width=1\linewidth]{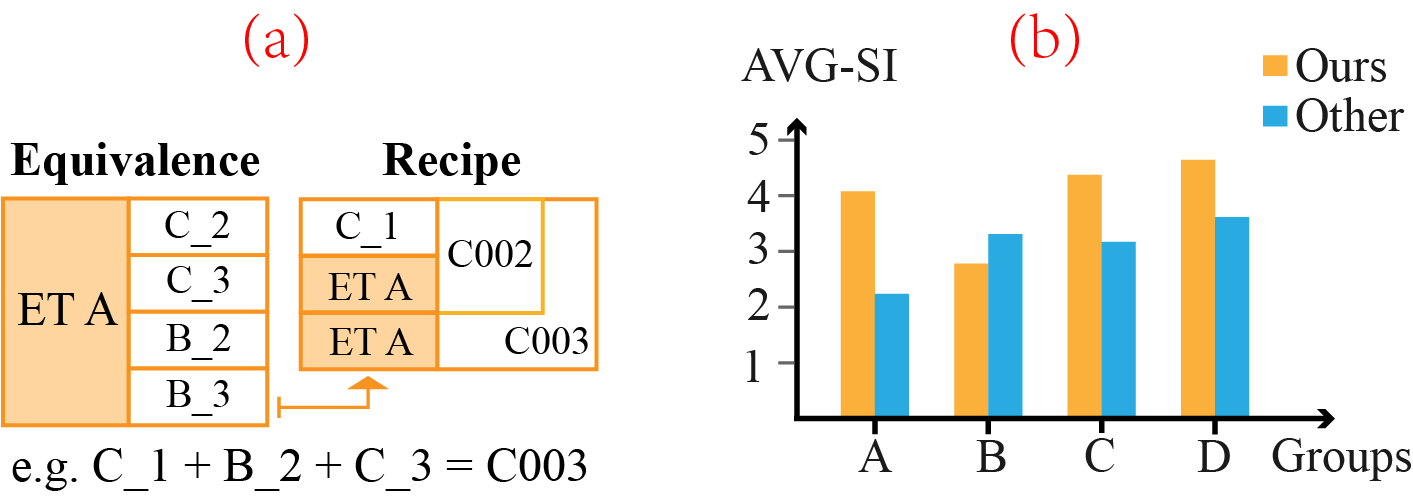}
    \caption{(a) Recipe table created based on the reusability of parts, (b) Results of user studies.}
    \label{fig:4}
\end{figure}

After the user enters the parts splicing area, he selects the appropriate parts, splices them into the corresponding Chinese characters and places them in the green area. The system captures the image of the area through the fixed-angle recognition camera (CaptureCamera) configured above the desktop and recognizes it. The recognition result will be compared with the core characters previously extracted and stored. If the match is successful, the user will get a round card for subsequent model activation. When the user picks up two parts to splice, the system will identify their numbers (Part ID) and check whether there are equivalent substitutes, and classify them into the same equivalent set (Equivalent Table). Then the recipe table is used to determine whether the two parts can be paired. If they match, the splicing is successful and a new part is generated, as shown in Figure \ref{fig:4}(a). After completing the parts assembly and successfully obtaining the round plate, the user can return to the model area and use the round plate to activate the previously generated model.

In the above stages,multiple users are supported to participate in the experience together,and each user can propose their own methods and suggestions to modify the data flow in each stage.

In this study, we recruit 16 participants to validate our system. Sixteen participants were divided into four groups. Each group was given the same test: first, they learned Chinese characters using the system, and then they learned the same characters using other methods. After the test, the participants rated the two learning methods based on four dimensions: immersion, convenience, fun, and information acquisition efficiency. Each dimension used a Likert scale of 1-5, with 1 indicating very dissatisfied and 5 indicating very satisfied. The average satisfaction of the testers (AVG-SI) was calculated for comparative analysis, as shown in Figure \ref{fig:4}(b).

\vspace{-1em}
\section{Result and Future Work}
Our HVRMT effectively combines Hanzi character acquisition with mortise-tenon joinery through VR. The mapping of strokes to components provides a tangible way to understand Hanzi structure.At the same time,the system's feature of allowing multiple people to participate together further enhances the interaction between users and improves the richness of the learning experience.Experimental results confirm the system’s effectiveness in improving memory retention and comprehension, with positive user feedback validating its technical feasibility and educational value. We plan to expand the content library with more Hanzi characters and mortise-tenon types. Further experiments with larger participant groups will assess the system’s impact on Hanzi literacy acquisition.

\bibliographystyle{splncs04}
\bibliography{refer}

\end{document}